\documentclass[final]{elsart}
\usepackage{graphicx}
\usepackage{amssymb,amsmath}
\usepackage[square,comma,numbers,sort&compress]{natbib}
\usepackage{amsfonts}
\usepackage{graphics}
\usepackage[thinspace,squaren]{mySIunits}

\def\etc{{\em etc.}}

\def\full{\protect\mbox{------}}
\def\kesik{\protect\mbox{-\, -\, -\, -}}

\begin{document}
\begin{frontmatter}
  \title{Elastic properties of highly anisotropic thin poly(propylene) foams}
  \author{Enis Tuncer\corauthref{cor}} \and
  \corauth[cor]{Corresponging author.}
  \ead{enis.tuncer@physics.org}
  \author{Michael Wegener}
%  \author{Reimund Gerhard-Multhaupt}
  \address{Applied Condensed-Matter Physics, Department of Physics, University of Potsdam, D-14469 Potsdam, Germany}

  \begin{abstract}
    In this letter, elastic properties of highly anisotropic cellular poly(propylene) films are reported. The material shows peculiar elastic properties compared to other foams in the literature. The data is displayed as the relative Young's modulus $E^*/E_s$ versus relative density $\rho^*/\rho_s$. Almost all the data from the literature are located on the region $E^*/E_s=(\rho^*/\rho_s)^n$ with $1\le n\le6$. The introduced material on the other hand have lower relative Young's modulus at high relative densities, $n\ge6$. %This characteristic is explained by the high anistropy in the material microstructure. % and verified by the computer simulations. %Such materials may be prefered when the material stores a soft materials 
  \end{abstract}
\end{frontmatter}
Foams are solid-gas mixtures, in which the solid phase forms a continuous (percolating) network structure, and performs an action depending on the application, {\it i.e.}, supports load, conducts heat/electricity,  {\it etc}.   The ratio of effective Young's modulus of foams to the solid's modulus $E^*/E_s$ has been related to the volume fraction of the solid material $\rho^*/\rho_s$~\cite{CellularSolids}.
  \begin{equation}
    \label{eq:power}
    E^*/E_s=C\,(\rho^*/\rho_s)^n
  \end{equation}
where $C$ and $n$ are constants and superscript `$*$' and and subscript `$s$' denote the properties of the effective and the solid material, respectively. Yet, there is no unique theory that explains the effective properties of foams or any composite that takes into account the geometrical parameters of the material. However, equations based on the effective medium theory and empirical mixture formulas have been proposed~\cite{TorquatoBook,MiltonBook}. For foams Eq.~(\ref{eq:power}) has been extensively used, and the constant $n$ often gives information about the structure of material in hand %, {\eg}, for closed and open cell structures $n\sim1$ and $n\sim2$  are observed 
\cite{CellularSolids,Garboczi2002a,Garboczi2002b,Garboczi2001,TuncerEyeTruss2003}.%, respectively. 

In Table~\ref{tab:info}, data taken from the literature for various foams and characteristic exponent values of Eq.~(\ref{eq:power}) are presented. In the table, closed cell foams have the lowest exponent $n$ which is around $\approx3/2$~\cite{Morgan1981,Zwissler1983,walsh1965,Mills1999,Baxter1972,Clutton1991,Scanlon2001,Christensen1986,Sun2002,Kitazono2003,Baltsavias1999,CellularSolids,Ladd1997,Hellmich2002}. Numerical simulations performed by considering three-dimensional closed-cell structures also yield similar exponents~\cite{Garboczi2002a,Garboczi2001}. In contrast to closed cell foams, most of the open cell foams result in $E^{*}/E_s\propto (\rho^*/\rho_s)^{2}$ dependence~\cite{CellularSolids,Hagiwara1987,Han2003,Badiche2000,Warren1988,Silva1997,Lederman1971,Gent1959,Sun2002,Garboczi2002b}. There are some exceptions which show $n>5/2$ dependence~\cite{Gross1997,Pekala1990,Lui2003,Badiche2000,Garboczi2002a,Fazekas2002}, which, we believe in principle, could be due to the anisotropy in materials. 

\begin{figure}[tp]
  \centering
  \includegraphics[width=5in]{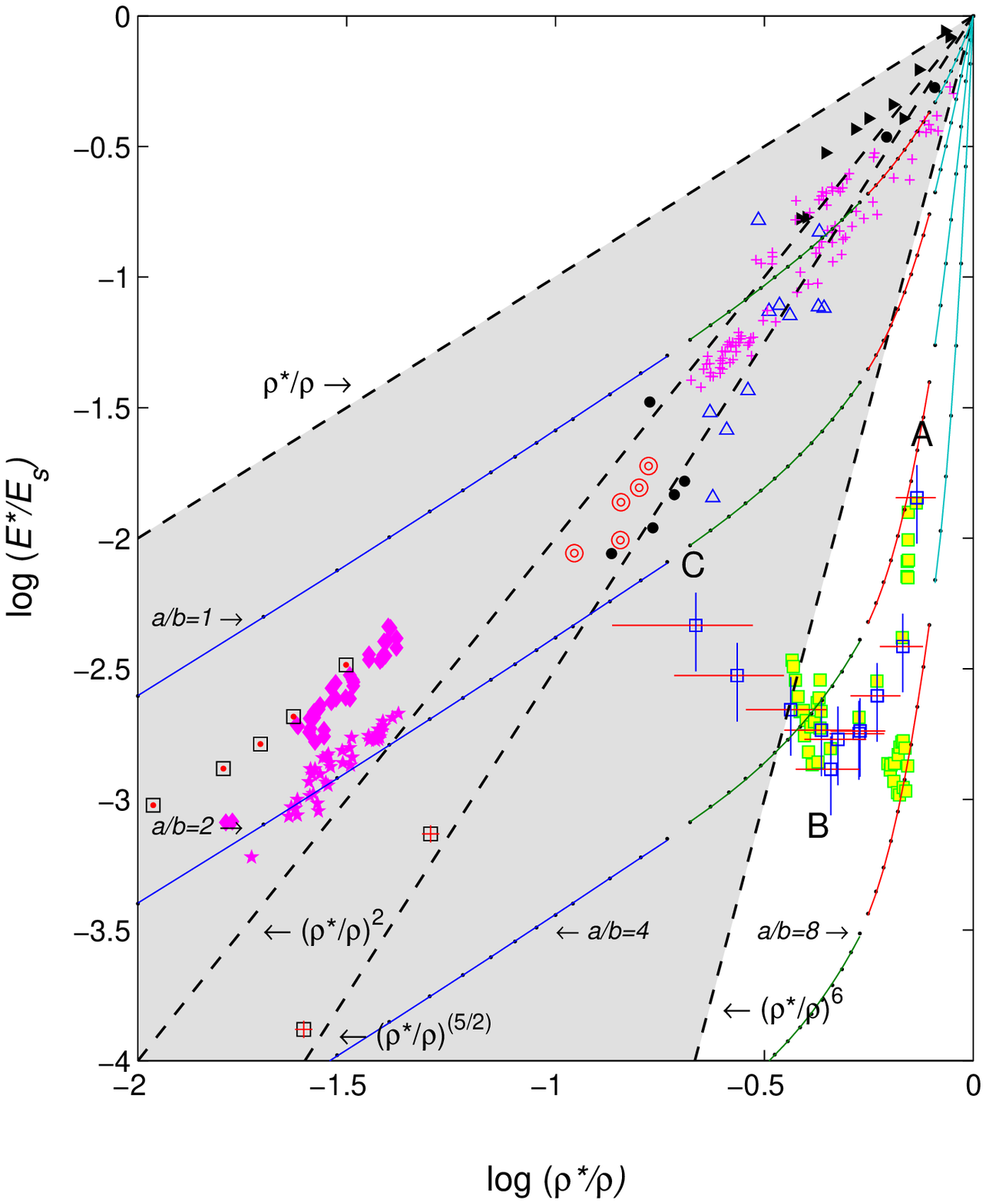}
  \caption{\label{fig:others} Relative effective Young's modulus versus relative density for different foams. The gray region marks the area of currently avaliable data present in the literature, Table~\ref{tab:info}. Some of the newly published data are shown with symbols; ($\ast$) \cite{Ladd1997}, ($\triangle$) \cite{Baltsavias1999}, ($\star$) and ($\blacklozenge$) \cite{Badiche2000}, ($\blacktriangleleft$) \cite{Sun2002}, ($+$) \cite{Hellmich2002}, ($\bullet$) \cite{Kitazono2003}, ($\boxplus$) \cite{jong1997}, ($\boxdot$) \cite{evans1997b} and ($\circledcirc$) \cite{ong2001}. The introduced and discussed poly(propylene) materials are plotted with ($\Box$) \cite{WegenerAPL2003} and ($\Box$) with errorbars.}
\end{figure}

The data from the given literature in Table~\ref{tab:info} are illustrated in Fig.~\ref{fig:others}. In the figure, the gray-region contains all the data presented in the literature (Table~\ref{tab:info}). Some of the newly published data are shown with symbols; ($\ast$) \cite{Ladd1997}, ($\triangle$) \cite{Baltsavias1999}, ($\star$) and ($\blacklozenge$) \cite{Badiche2000}, ($\blacktriangleleft$) \cite{Sun2002}, ($+$) \cite{Hellmich2002}, ($\bullet$) \cite{Kitazono2003}, ($\boxplus$) \cite{jong1997}, ($\boxdot$) \cite{evans1997b} and ($\circledcirc$) \cite{ong2001}. The data presented in this study are poly(propylene) foams and are plotted with ($\Box$); our data with errorbars and other performed experiments \cite{WegenerAPL2003}. The dashed lines (\kesik) shows Eq.~(\ref{eq:power}) for $C=1$ and $n=\{1,\,2,\,5/2,\,6\}$. All the data from the literature are located between $n\approx1$ and $n\approx 6$ region. The solid lines (\full) in the figure are calculated for a truss-like structure using the finite element method \cite{TuncerWegener2003}, which has voids as in the form of lozenge, with various void sizes, $a/b=\{1,\,2,\,4,\,8\}$. The parameters $a$ and $b$ define the void dimensions in the $x$- and $y$-dimensions, respectively--when the ratio $a/b$ is large the structure is highly anisotropic. At high concentrations of solid, the computer model based on the finite element method yield similar values as the experiments for $4<a/b<8$. 

The original films in this study were cellular poly(propylene) with typical thicknesses around $30-100\ \micro\meter$. They were stretched by the manufacturer (VTT Processes, Finland)\cite{PaajanenSensors2000} and the lateral dimensions were much larger than the thickness of the samples. The structure of samples resembles the puff pastry, some scanning electron micrographs have been presented in Ref.~\cite{TuncerWegener2003,WegenerPhysD2004,WegenerAPL2003,Wegener-CEIDP03}. The surface area of samples are larger than $1\ \centi\meter^2$  for our electromechanical applications. The average void structure is oblate with dimensions around $\sim50\ \micro\meter$ long and $\sim10\ \micro\meter$ high from the scanning electron micro-graphs, which indicates that there are at least $5$ voids in the thickness direction.  The original films were inflated with a post-processing procedure \cite{WegenerAPL2003}. Their elastic properties were measured after each inflation process with the dielectric spectroscopy method~\cite{AxelReview}. The inflation influences the relative density and average void (cell) dimension of the films. The change in the relative Young's moduli of the films is illustrated in the figure with open symbols. The relative Young's modulus and density were calculated by using $E_s=1\ \giga\pascal$ and $\rho_s=1\ \gram/\centi\meter^{3}$ for the solid poly(propylene). The letters {\sf A}, {\sf B} and {\sf C} present the different stages in the inflation process. The non-inflated films had the highest Young's moduli ({\sf A}). By inflating the samples, the stiffness of the films lowered down to $E^*/E_s\approx10^{-3}$ ({\sf B}). After this point, a further inflation of the films resulted an increase in the stiffness ({\sf C}). This was due to the shape of the voids, which became nearly-isotropic. The voids become rounder with $a/b\le4$--the anisotropy in the films were destroyed by the inflation.  The numerical simulations utilizing the finite element method by means of simple truss-like structure \cite{TuncerWegener2003} has showed a similar behavior (even simulations with more realistic geometries yield similar results \cite{TuncerEyeTruss2003}), and has verified the experimental observations. The `{\sf U}'-shaped behavior of our data could be explained by the solid materials' cell-wall thickness \cite{TuncerWegener2003,TuncerEyeTruss2003}. It should be mentioned that if the layered texture is exchanged to a fibrous one, one might obtain much softer materials with exponents $n>6$ in Eq.~(\ref{eq:power}). %One might also argue that the peculiar behavior of the present materials are due to their somewhat two-dimensional structure, physical strucuture changes only in two direction.

Almost all of our measurement points (only two of the data points fall in the region marked by $1<n<6$) are outside the region for the conventional foams. So we can conclude that as the films become more isotropic ($a/b\le2$),  they yield similar Young's modulus behavior as the foams presented in the literature, $n\le6$ in Eq.~(\ref{eq:power}). Although in the literature it is not clearly stated that foams with higher characteristic exponents in Eq.~(\ref{eq:power}) are anistropic, our observations affirm that anisotropic foams yield high exponent values for the power-law expression in Eq.~(\ref{eq:power}).

Finally, as a concluding remark, although poly(propylene) foam is widely used and has been known for decades, the presented low stiffness character of highly anistropic materials at high relative densities can be advantages and preferably selected for special applications in which a multifunctional material with soft elastic properties is needed. The solid phase in that case is utilized to possess an additional feature, such as storing charge, heat, magnetic/electric polarization \etc\ A discrete example is electret materials used in transducer applications~\cite{ReimundRev,ElectretPhysicsToday}. The solid phase in cellular electrets is used to store electrical charge, for this reason, in order to have high charge densities, high volume fraction of solid is simultaneously desired  together with low elastic constants. In addition, one can even tailor the elastic properties by using post processing techniques to change the micro-structure of the material. 

\begin{table}[tp]
  \caption{Effective Young's modulus relative density relation for different structures, Eq.~(\ref{eq:power}).\label{tab:info}}
  \centering
  \begin{tabular*}{6in}{l@{\extracolsep{\fill}}ll}
\hline\hline
 $1 < n < 2$ &   $3/2<n< 5/2$ &$n>5/2$ \\
\hline
 Closed-cell & Open-cells & Anisotropic-cells(?)\\
\hline
 Foamed glasses \citep{Morgan1981,Zwissler1983,walsh1965} &  Metal foams~\citep{CellularSolids,Hagiwara1987,Han2003,Badiche2000,Warren1988,Silva1997} & Aerogels~\citep{Gross1997,Pekala1990}\\
  Polymer foams~\citep{Christensen1986,Sun2002,Mills1999,Baxter1972,Clutton1991} & Polymer foams~\citep{Lederman1971,Gent1959,Sun2002,jong1997,evans1997b} & Honeycomb\citep{Badiche2000}\\
Metal foams~\citep{Kitazono2003,ong2001} & Numerical~\citep{Garboczi2002b} &Penrose network\citep{Badiche2000}\\
Composite food~\citep{Baltsavias1999,Scanlon2001} & & Various foams\citep{Lui2003}\\
Biological materials~\citep{CellularSolids,Ladd1997,Hellmich2002} & &Numerical~\citep{Garboczi2002a,Fazekas2002} \\
Numerical~\citep{Garboczi2002a,Garboczi2001} & &\\
\hline\hline
%& \citep{CellularSolids} & \citep{Badiche2000} &\citep{Warren1988} &\citep{Badiche2000}\\
  \end{tabular*}  
\end{table}

%\bibliography{../../newref.bib}
%\bibliographystyle{phaip}
%\bibliographystyle{elsart-num}
\bibliographystyle{unsrtnat}

\end{document}